\documentclass[12pt,aps,onecolumn,superscriptaddress]{revtex4}
\usepackage{graphicx}
\usepackage{epsfig}
\usepackage{amssymb,amsmath}

\begin{document}

\title{\large\bf\boldmath 
Search for the process $e^+e^-\to f_1(1285)$ at the SND detector
}

\author{M.~N.~Achasov} 
\affiliation{Budker Institute of Nuclear Physics, SB RAS, Novosibirsk, 630090, Russia} 
\affiliation{Novosibirsk State University, Novosibirsk, 630090, Russia} 
\author{A.~E.~Alizzi}
\affiliation{Budker Institute of Nuclear Physics, SB RAS, Novosibirsk, 630090, Russia}
\affiliation{Novosibirsk State University, Novosibirsk, 630090, Russia}
\author{A.~Yu.~Barnyakov} 
\affiliation{Budker Institute of Nuclear Physics, SB RAS, Novosibirsk, 630090, Russia}
\author{E.~V.~Bedarev}
\affiliation{Budker Institute of Nuclear Physics, SB RAS, Novosibirsk, 630090, Russia} 
\author{K.~I.~Beloborodov}
\affiliation{Budker Institute of Nuclear Physics, SB RAS, Novosibirsk, 630090, Russia} 
\affiliation{Novosibirsk State University, Novosibirsk, 630090, Russia} 
\author{A.~V.~Berdyugin} 
\affiliation{Budker Institute of Nuclear Physics, SB RAS, Novosibirsk, 630090, Russia} 
\affiliation{Novosibirsk State University, Novosibirsk, 630090, Russia} 
\author{D.~E.~Berkaev} 
\affiliation{Budker Institute of Nuclear Physics, SB RAS, Novosibirsk, 630090, Russia} 
\affiliation{Novosibirsk State University, Novosibirsk, 630090, Russia} 
\author{A.~G.~Bogdanchikov} 
\affiliation{Budker Institute of Nuclear Physics, SB RAS, Novosibirsk, 630090, Russia} 
\author{A.~A.~Botov} 
\affiliation{Budker Institute of Nuclear Physics, SB RAS, Novosibirsk, 630090, Russia} 
\author{D.~E.~Chistyakov} 
\affiliation{Budker Institute of Nuclear Physics, SB RAS, Novosibirsk, 630090, Russia} 
\author{T.~V.~Dimova} 
\affiliation{Budker Institute of Nuclear Physics, SB RAS, Novosibirsk, 630090, Russia} 
\affiliation{Novosibirsk State University, Novosibirsk, 630090, Russia} 
\author{V.~P.~Druzhinin} 
\email{druzhinin@inp.nsk.su}
\affiliation{Budker Institute of Nuclear Physics, SB RAS, Novosibirsk, 630090, Russia} 
\affiliation{Novosibirsk State University, Novosibirsk, 630090, Russia} 
\author{R.~A.~Efremov}
\affiliation{Budker Institute of Nuclear Physics, SB RAS, Novosibirsk, 630090, Russia} 
\author{L.~V.~Kardapoltsev}
\affiliation{Budker Institute of Nuclear Physics, SB RAS, Novosibirsk, 630090, Russia} 
\affiliation{Novosibirsk State University, Novosibirsk, 630090, Russia} 
\author{A.~S.~Kasaev}
\affiliation{Budker Institute of Nuclear Physics, SB RAS, Novosibirsk, 630090, Russia} 
\author{A.~A.~Kattsin}
\affiliation{Budker Institute of Nuclear Physics, SB RAS, Novosibirsk, 630090, Russia} 
\author{V.~R.~Khamidullin} 
\affiliation{Budker Institute of Nuclear Physics, SB RAS, Novosibirsk, 630090, Russia} 
\affiliation{Novosibirsk State University, Novosibirsk, 630090, Russia} 
\author{A.~G.~Kharlamov} 
\affiliation{Budker Institute of Nuclear Physics, SB RAS, Novosibirsk, 630090, Russia} 
\affiliation{Novosibirsk State University, Novosibirsk, 630090, Russia} 
\author{I.~A.~Koop}
\affiliation{Budker Institute of Nuclear Physics, SB RAS, Novosibirsk, 630090, Russia} 
\affiliation{Novosibirsk State University, Novosibirsk, 630090, Russia} 
\author{A.~A.~Korol} 
\affiliation{Budker Institute of Nuclear Physics, SB RAS, Novosibirsk, 630090, Russia} 
\affiliation{Novosibirsk State University, Novosibirsk, 630090, Russia} 
\author{D.~P.~Kovrizhin} 
\affiliation{Budker Institute of Nuclear Physics, SB RAS, Novosibirsk, 630090, Russia} 
\author{A.~S.~Kupich} 
\affiliation{Budker Institute of Nuclear Physics, SB RAS, Novosibirsk, 630090, Russia} 
\affiliation{Novosibirsk State University, Novosibirsk, 630090, Russia} 
\author{A.~P.~Kryukov} 
\affiliation{Budker Institute of Nuclear Physics, SB RAS, Novosibirsk, 630090, Russia} 
\author{A.~R.~Mamedov} 
\affiliation{Budker Institute of Nuclear Physics, SB RAS, Novosibirsk, 630090, Russia} 
\affiliation{Novosibirsk State University, Novosibirsk, 630090, Russia} 
\author{N.~A.~Melnikova} 
\affiliation{Budker Institute of Nuclear Physics, SB RAS, Novosibirsk, 630090, Russia} 
\author{N.~Yu.~Muchnoi} 
\affiliation{Budker Institute of Nuclear Physics, SB RAS, Novosibirsk, 630090, Russia} 
\affiliation{Novosibirsk State University, Novosibirsk, 630090, Russia} 
\author{A.~E.~Obrazovsky} 
\affiliation{Budker Institute of Nuclear Physics, SB RAS, Novosibirsk, 630090, Russia} 
\author{A.~A.~Oorzhak}
\affiliation{Budker Institute of Nuclear Physics, SB RAS, Novosibirsk, 630090, Russia}
\affiliation{Novosibirsk State University, Novosibirsk, 630090, Russia}
\author{I.~V.~Ovtin}
\affiliation{Budker Institute of Nuclear Physics, SB RAS, Novosibirsk, 630090, Russia}
\affiliation{Novosibirsk State University, Novosibirsk, 630090, Russia}
\author{E.~V.~Pakhtusova} 
\affiliation{Budker Institute of Nuclear Physics, SB RAS, Novosibirsk, 630090, Russia} 
\author{I.~A.~Polomoshnov}
\affiliation{Budker Institute of Nuclear Physics, SB RAS, Novosibirsk, 630090, Russia}
\affiliation{Novosibirsk State University, Novosibirsk, 630090, Russia}
\author{K.~V.~Pugachev} 
\affiliation{Budker Institute of Nuclear Physics, SB RAS, Novosibirsk, 630090, Russia} 
\affiliation{Novosibirsk State University, Novosibirsk, 630090, Russia} 
\author{S.~A.~Rastigeev} 
\affiliation{Budker Institute of Nuclear Physics, SB RAS, Novosibirsk, 630090, Russia} 
\author{Yu.~A.~Rogovsky} 
\affiliation{Budker Institute of Nuclear Physics, SB RAS, Novosibirsk, 630090, Russia} 
\affiliation{Novosibirsk State University, Novosibirsk, 630090, Russia} 
\author{V.~A.~Romanov} 
\affiliation{Budker Institute of Nuclear Physics, SB RAS, Novosibirsk, 630090, Russia} 
\affiliation{Novosibirsk State University, Novosibirsk, 630090, Russia} 
\author{S.~I.~Serednyakov} 
\affiliation{Budker Institute of Nuclear Physics, SB RAS, Novosibirsk, 630090, Russia} 
\affiliation{Novosibirsk State University, Novosibirsk, 630090, Russia} 
\author{D.~A.~Shtol} 
\affiliation{Budker Institute of Nuclear Physics, SB RAS, Novosibirsk, 630090, Russia} 
\author{Z.~K.~Silagadze} 
\affiliation{Budker Institute of Nuclear Physics, SB RAS, Novosibirsk, 630090, Russia} 
\affiliation{Novosibirsk State University, Novosibirsk, 630090, Russia} 
\author{K.~D.~Sungurov}
\affiliation{Budker Institute of Nuclear Physics, SB RAS, Novosibirsk, 630090, Russia} 
\affiliation{Novosibirsk State University, Novosibirsk, 630090, Russia} 
\author{I.~K.~Surin} 
\affiliation{Budker Institute of Nuclear Physics, SB RAS, Novosibirsk, 630090, Russia} 
\author{Yu.~V.~Usov} 
\affiliation{Budker Institute of Nuclear Physics, SB RAS, Novosibirsk, 630090, Russia} 
\author{V.~N.~Zhabin} 
\affiliation{Budker Institute of Nuclear Physics, SB RAS, Novosibirsk, 630090, Russia} 
\affiliation{Novosibirsk State University, Novosibirsk, 630090, Russia} 
\author{V.~V.~Zhulanov}
\affiliation{Budker Institute of Nuclear Physics, SB RAS, Novosibirsk, 630090, Russia}
\affiliation{Novosibirsk State University, Novosibirsk, 630090, Russia}

\begin{abstract}
In the experiment with the SND detector at the VEPP-2000 $e^+e^-$
collider, a search is performed for the direct production of the
$C$-even $f_1(1285)$ resonance in $e^+e^-$ collisions. The analysis is
based on data with an integrated luminosity of about 200 pb$^{-1}$,
accumulated in the center-of-mass energy range of 1.14--1.46 GeV, of which
about 72 pb$^{-1}$ were recorded near the maximum of the $f_1(1285)$ resonance.
The $f_1(1285)$ production cross section at the resonance maximum
$\sigma(e^+e^-\to f_1)=(31\pm 13\pm 4)$ pb and the branching fraction
$B(f_1(1285)\to e^+e^-)=(3.5\pm 1.4\pm 0.5)\times 10^{-9}$ have been measured.
The significance of the $e^+e^-\to f_1(1285)$
signal is $2.4\sigma$. Since the significance is low, we also present the
upper limits at the 90\% confidence level: 
$\sigma(e^+e^-\to f_1)<48\mbox{ pb}$
and $B(f_1(1285)\to e^+e^-)<5.4\times 10^{-9}$.
\end{abstract}

\maketitle

\section{Introduction}
The dominant mechanism of hadron production in $e^+e^-$ collisions is
single-photon annihilation. In particular, for the
production of a single resonance $e^+e^-\to R$, its quantum numbers
coincide with the quantum numbers of the photon $J^{PC}=1^{--}$.
Annihilation through two photons is suppressed by a factor of $\alpha^2$,
where $\alpha$ is the fine structure constant. In
two-photon annihilation, $C$-even resonances are produced. Experiments
to search for the production of $C$-even resonances began more than 30
years ago at the VEPP-2M collider with the ND detector~\cite{ND}. At the
VEPP-2M~\cite{ND,SND1}, VEPP-2000~\cite{SND2,SND3} and
BEPCII~\cite{bes} colliders, upper limits were set on the production
of $\eta$, $\eta^\prime$, $f_0(975)$, $f_2(1270)$, $f_1(1285)$, $f_0(1300)$,
$a_0(980)$, $a_2(1320)$, and $X(3872)$. The only observed process of
direct production of the $C$-even resonance was $e^+e^-\to\chi_{c1}$,
in the BESIII experiment~\cite{beschic1}.

\begin{figure}[htbp]
\includegraphics[width=0.65\textwidth]{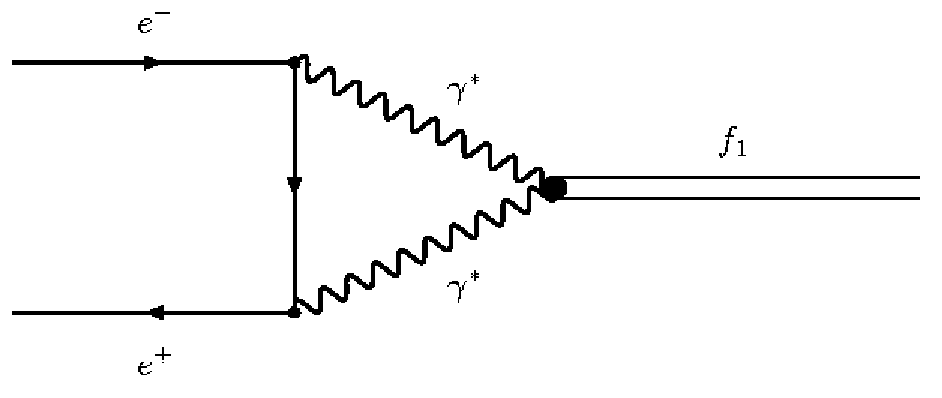}
\caption{ \label{fig1}
The diagram for the process $e^+e^-\to f_1$.}
\end{figure}

In this paper, we search for the process $e^+e^-\to f_1(1285)$, the
diagram for which is shown in Fig.~\ref{fig1}. Axial-vector resonances
do not decay into two photons, but can nevertheless be produced in
$e^+e^-$ annihilation through virtual photons. The cross section at
the resonance maximum is related to its leptonic branching fraction as
follows: $\sigma(e^+e^-\to f_1)=(12\pi/m^2_{f_1})B(f_1\to e^+e^-)$. 
The measurement of the decay $f_1\to e^+e^-$ is important for testing
models of $f_1$ transition form factors~\cite{ZBK}. The form factors of 
axial-vector mesons are 
used to calculate their contribution to hadronic light-by-light scattering 
in the anomalous magnetic moment of the muon. 

In Refs.~\cite{Rudenko,HKZ},
the transition form factors are studied within the vector-meson dominance
model. The form factor parameters are determined from experimental
data on $f_1$ meson decays and the process $\gamma\gamma^\ast\to f_1$.
In Ref.~\cite{HKZ}, data on the process $e^+e^-\to f_1\pi^+\pi^-$
are additionally taken into account. The obtained form factors are used to
predict $B(f_1\to e^+e^-)=(3.5\pm1.8)\times10^{-9}$~\cite{Rudenko}
and $(2.2\pm0.6)\times10^{-9}$~\cite{HKZ}. 
The corresponding cross sections $\sigma(e^+e^-\to f_1)$ are 
$31\pm16$ pb and  $20\pm5$ pb.

The main decay modes of the $f_1$ meson are $\pi^+\pi^-\pi^0\pi^0$,
$\pi^+\pi^-\pi^+\pi^-$, $\eta\pi^+\pi^-$, and $\eta\pi^0\pi^0$. The
first three final states are produced in single-photon annihilation
and have cross sections at $\sqrt{s}=m_{f_1}$ that are 2--3 orders of
magnitude larger than the prediction for $\sigma(e^+e^-\to f_1)$.
Therefore, the best mode to search for the process $e^+e^-\to
f_1(1285)$ is $f_1\to\eta\pi^0\pi^0$, whose branching fraction is
$(17.3\pm0.7)\%$~\cite{pdg}.

The results of a search for direct production of the $f_1(1285)$
meson, using a part of the statistics accumulated in the SND experiment at the
VEPP-2000 $e^+e^-$ collider~\cite{VEPP2000}, were published in 
Ref.~\cite{SNDf1}. That study analyzed data corresponding to an integrated 
luminosity of 15.1 pb$^{-1}$, recorded over the center-of-mass (c.m.) energy 
range $E\equiv\sqrt{s}=1.2\mbox{--}1.4$ GeV, at 12 energy points,
with 3.4 pb$^{-1}$ collected at the $f_1$ resonance peak.
Only two events at the resonance peak were selected for analysis.
Unfortunately, a technical mistake was introduced when fitting
the energy distribution of events by the sum of the signal and
background distributions, which led to an overestimate of the
significance of the $f_1$ signal. We refit the data presented
in Table II of Ref.~\cite{SNDf1} and obtain that the significance of the
$f_1$ signal is $1.8\sigma$. The uncertainties in the
measured cross section and branching fraction increase accordingly:
\begin{equation}
\sigma(e^+e^-\to f_1)=45^{+49}_{-31}\mbox{ pb, }
B(f_1\to e^+e^-)=5.1^{+5.6}_{-3.5}\times 10^{-9},
\label{resultold}
\end{equation} 
and the upper limit at the 90\% confidence level is
\begin{equation}
B(f_1\to e^+e^-)<11.6\times 10^{-9}.
\label{limitold}
\end{equation}

Recently, a preprint on a search for the $f_1\to e^+e^-$ process in
the CMD-3 experiment at the VEPP-2000 collider was
published~\cite{cmdf1}. The CMD-3 result, which was also obtained in the
$f_1\to\eta\pi^0\pi^0$ decay mode, is $B(f_1\to e^+e^-)=(8.3\pm 2.3\pm
1.1)\times 10^{-9}$.

\section{Detector and experiment}
In this paper, we analyze data with an integrated luminosity of about
200 pb$^{-1}$, recorded in 2017, 2019, 2023, and 2024 in the energy
range $\sqrt{s}=1.14\mbox{--}1.46$ GeV at 23 energy points, with
about 72 pb$^{-1}$ collected at two points (1.276 and 1.282 GeV) near the
$f_1(1285)$ resonance maximum.

A detailed description of the SND detector is given in Ref.~\cite{SND}.
It is a non-magnetic detector, the main part of
which is a three-layer spherical electromagnetic calorimeter based on
NaI(Tl) crystals. The solid angle of the calorimeter is 95\% of
4$\pi$. Its energy resolution for photons is
$\sigma_{E_\gamma}/E_\gamma=4.2\%/\sqrt[4]{E_\gamma({\rm GeV})}$, and
the angular resolution is about $1.5^\circ$. The directions of charged
particles are measured in a tracking system consisting of a nine-layer
drift chamber and a proportional chamber with cathode-strip readout.
The solid angle of the tracking system is 94\% of 4$\pi$. The
calorimeter is surrounded by a muon system consisting of proportional
tubes and scintillation counters. 

The search for the $e^+e^-\to f_1(1285)$ process is carried out in the
$f_1(1285)\to\eta\pi^0\pi^0$ channel, followed by the decay of the
$\eta$ and $\pi^0$ mesons into two photons. Since the final state for
the process under study does not contain charged particles, the
process without charged particles, $e^+e^-\to \gamma\gamma$,  is chosen
for normalization. This normalization reduces the systematic
uncertainties associated with the event selection in the hardware
first-level trigger, as well as the uncertainties arising from
superimposing beam-generated spurious tracks and photons onto the
events under study. The accuracy of luminosity measurement with the
$e^+e^-\to \gamma\gamma$ process is no worse than 1\%. The distribution of the
integrated luminosity over 23 energy points is listed in Table~\ref{tab2}.

According to the Particle Data Group (PDG)~\cite{pdg}, the dominant
intermediate state in the $f_1(1285)\to\eta\pi\pi$ decay is $a_0(980)\pi$.
Its fraction is $(72\pm8)\%$. The process $e^+e^- \to f_1(1285) \to
a_0^0\pi^0 \to \eta\pi^0\pi^0$ is modeled using the formulas from
Ref.~\cite{Rudenko}. For the remaining 28\% of the
$f_1(1285)\to\eta\pi\pi$ decay, the intermediate state
$f_0(500)\eta$ is used.

In the energy range under study, the background sources are the
following processes with multiphoton final states:
$e^+e^-\to \pi^0\pi^0\gamma$,
$e^+e^-\to\omega\pi^0\pi^0\to 3\pi^0\gamma$,
$e^+e^-\to\eta\gamma(\gamma)\to 3\pi^0\gamma(\gamma)$, 
$e^+e^-\to K_SK_L(\gamma)\to 2\pi^0 K_L(\gamma)$, and
$e^+e^-\to\eta\pi^0\gamma$.
In the $e^+e^-\to \pi^0\pi^0\gamma$ process, the intermediate state
$\omega\pi^0$ dominates. The contribution of
non-$\omega\pi^0$ mechanisms is about 1\%. This contribution is 
simulated with a uniform phase-space distribution. For
the $e^+e^-\to\eta\pi^0\gamma$ process, there are also two
mechanisms: purely hadronic ($\omega\eta$) and radiative~\cite{SNDetpg}.
The radiative mechanism is especially important in this energy range
and is simulated using a uniform phase-space distribution.

The simulation of the signal and background processes includes radiative
corrections~\cite{rad1}. The angular distribution of the photon emitted
from the initial state is generated according to Ref.~\cite{rad2}. 
The Born cross sections used in the simulation are taken from
Ref.~\cite{SNDppg} for $e^+e^-\to\pi^0\pi^0\gamma$, 
Refs.~\cite{ompipi1,ompipi2} for $e^+e^-\to\omega\pi\pi$,
Refs.~\cite{etag1,etag2} for $e^+e^-\to\eta\gamma$, 
Refs.~\cite{SNDkskl,kskl1} for $e^+e^-\to K_SK_L$,
and Ref.~\cite{SNDetpg} for $e^+e^-\to \eta\pi^0\gamma$. For the $e^+e^-\to
\eta\pi^0\gamma$ process, the Born cross section for the radiative mechanism is
replaced by a new one obtained from the full SND data set.

In the processes $e^+e^-\to\pi^0\pi^0\gamma$ and
$e^+e^-\to\eta\pi^0\gamma$, the extra photons appear due to
splitting of the electromagnetic showers, photon emission by the
initial particles at a large angle, and superimposing beam-generated
background.  To simulate the latter effect, background events recorded
during the experiment with a special random trigger are mixed with the
simulated events. 

For the processes $e^+e^-\to\eta\gamma(\gamma)$ and $e^+e^-\to
K_SK_L(\gamma)$, the photon in parentheses corresponds to
initial-state radiation. Since the Born cross sections for these
processes are small in the energy range under study, the main contribution to
the background comes from the radiative return to the $\phi(1020)$ resonance,
$e^+e^-\to \phi(1020)\gamma$.

\section{Selection criteria\label{sel}}
For the search for the process $e^+e^- \to f_1(1285) \to
\eta\pi^0\pi^0$, we use selection criteria close to those in
the previous SND analysis~\cite{SNDf1}. Events with exactly six 
reconstructed photons and no tracks
in the drift chamber are selected. Photons are clusters in the
calorimeter with an energy deposition greater than 20 MeV. The total
energy deposition in the calorimeter $E_{tot}$ and the total momentum
of the event $P_{tot}$, calculated from the energy depositions in the
calorimeter crystals, must satisfy the following conditions:
\begin{equation}
E_{tot}/\sqrt{s} > 0.6,~P_{tot}/\sqrt{s} < 0.3.
\label{eton_vs_ptrt}
\end{equation}

\begin{figure}
\includegraphics[width=0.47\textwidth]{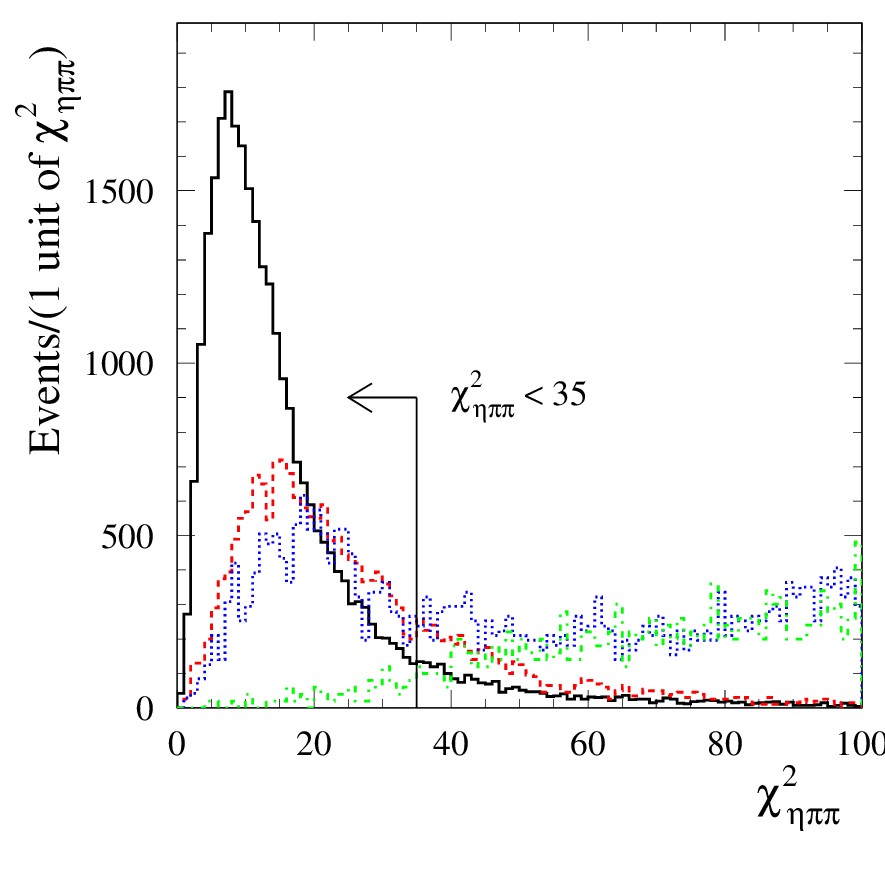}
\includegraphics[width=0.47\textwidth]{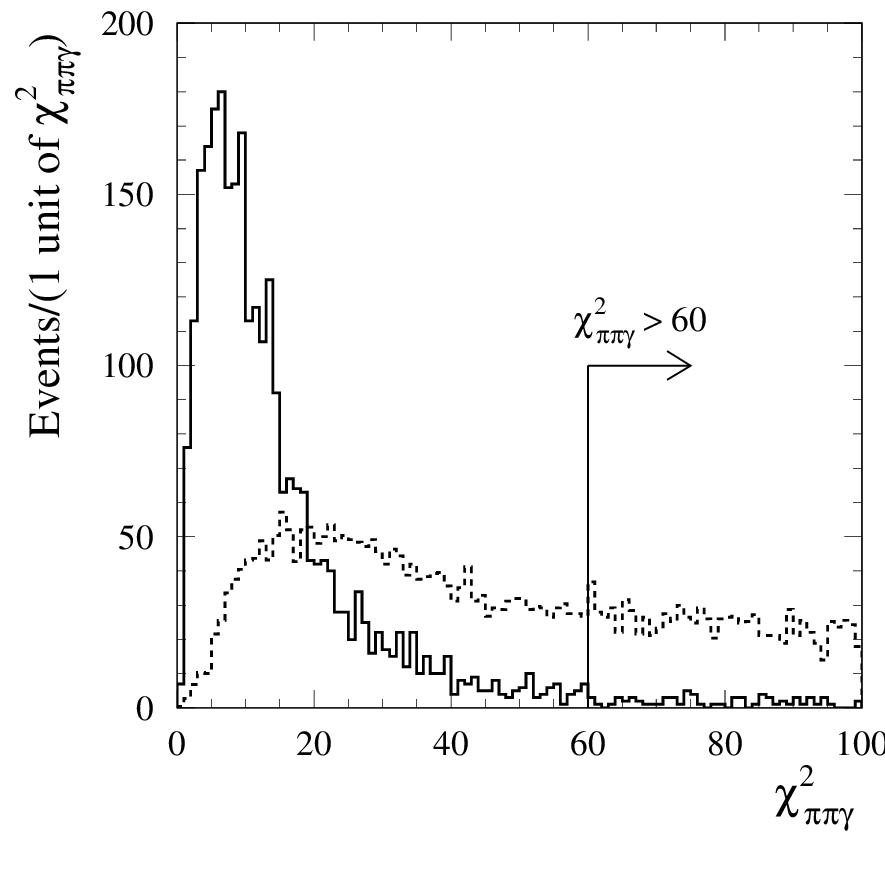}
\caption{Left panel: The $\chi^2_{\eta\pi\pi}$ distribution for
simulated events of the process under study $e^+e^-\to\eta\pi^0\pi^0$
(solid black histogram) and background processes $e^+e^-\to\pi^0\pi^0\gamma$
(dashed red histogram), $e^+e^-\to\eta\gamma$ (dotted blue histogram),
and $e^+e^-\to K_SK_L$ (dash-dotted green histogram) at
$\sqrt{s}=1282$ MeV. The arrow indicates the cut
$\chi^2_{\eta\pi\pi}<35$.
Right panel: The $\chi^2_{\pi\pi\gamma}$  distribution for simulated
events of the processes $e^+e^-\to \omega\pi^0(\gamma)$ (solid histogram)
and $e^+e^-\to f_1 \to \eta\pi^0\pi^0$ (dashed histogram) at
$\sqrt{s}=1282$ MeV. The arrow
indicates the cut $\chi^2_{\pi\pi\gamma}>60$.
\label{fig2}}
\end{figure}

For the selected events, a kinematic fit is performed to the
hypothesis $e^+e^-\to 6\gamma$. Then, using the refined photon
parameters, all possible two-photon invariant masses are calculated.
Events are required to contain two pairs of photons with an
invariant mass in the range $m_{\pi^0}\pm 35$ MeV and one pair with an
invariant mass in the range $m_{\eta}\pm 50$ MeV. 
For such events, a kinematic fit is performed to the hypothesis
$e^+e^-\to\eta\pi^0\pi^0\to 6\gamma$. The distribution of the kinematic fit
$\chi^2$ ($\chi^2_{\eta\pi\pi}$) for simulated events of the signal and 
selected background processes is shown in Fig.~\ref{fig2} (left).
The condition $\chi^2_{\eta\pi\pi}<35$ is imposed.
It is evident that this condition removes a significant fraction of
events from the $e^+e^-\to\eta\gamma$ and $e^+e^-\to K_SK_L$ processes.
However, for the dominant background process
$e^+e^-\to\pi^0\pi^0\gamma$, the suppression is small.
\begin{figure}
\includegraphics[width=0.47\textwidth]{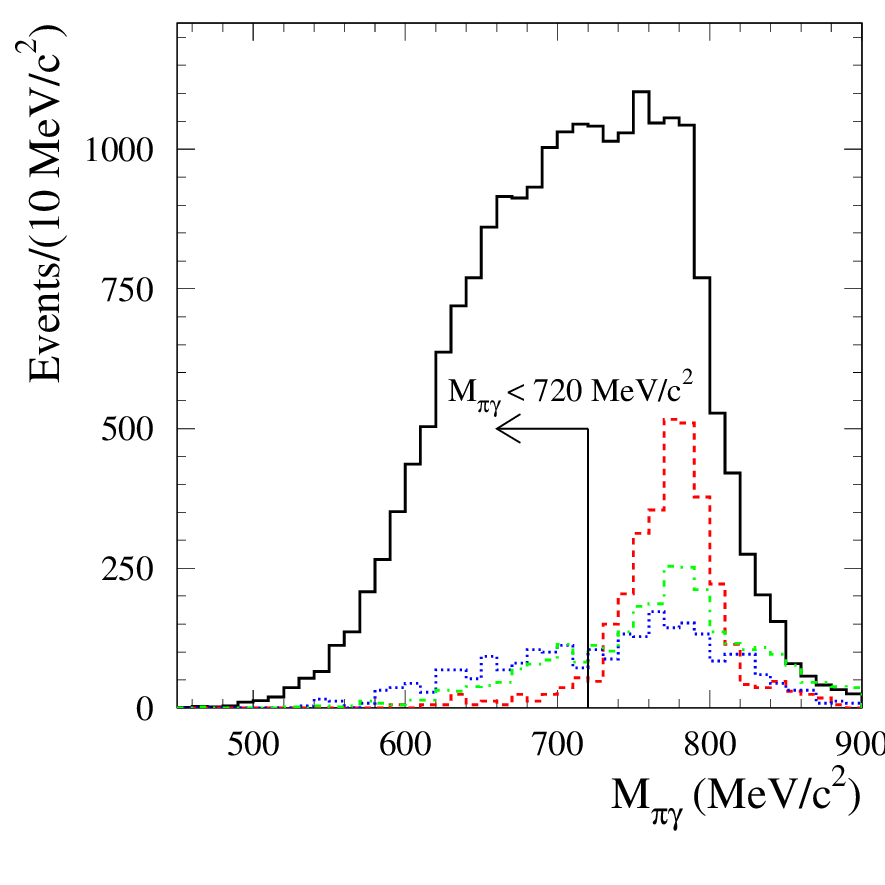}
\includegraphics[width=0.47\textwidth]{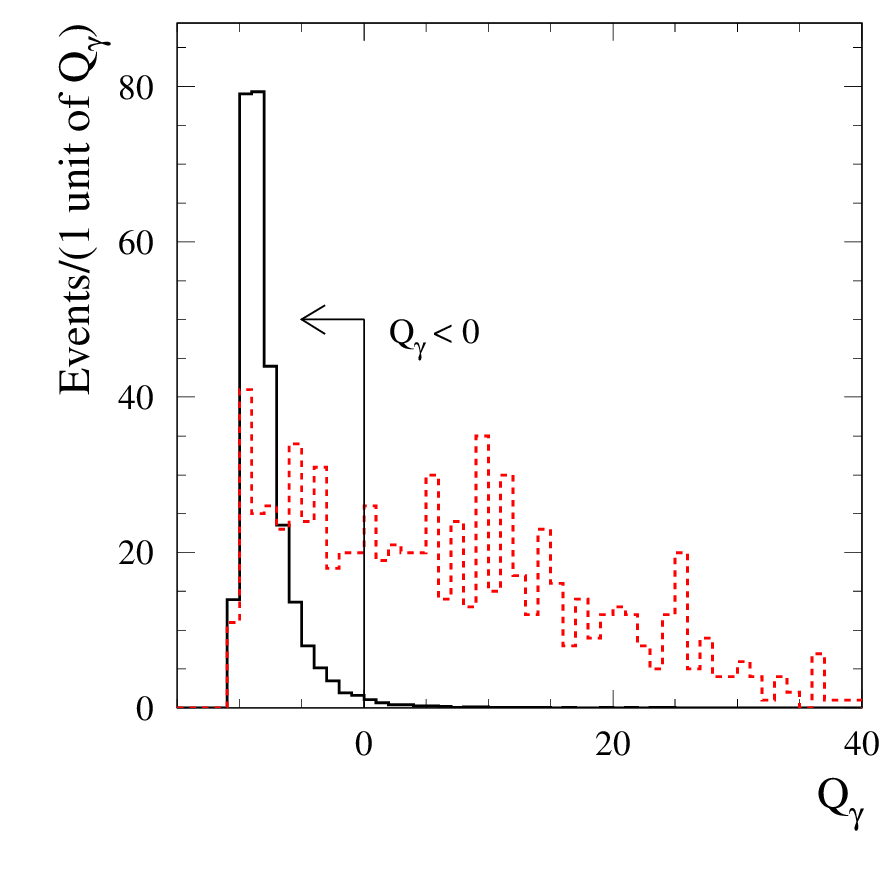}
\caption{Left panel: The distribution of the invariant mass
$M_{\pi\gamma}$ of the $\omega$-meson candidate for simulated events
of the processes $e^+e^-\to f_1 \to \eta\pi^0\pi^0$ (solid black
histogram), $e^+e^-\to\omega\pi^0\to\pi^0\pi^0\gamma$ (dashed red
histogram), $e^+e^-\to\pi^0\pi^0\gamma$ with the uniform phase-space
distribution (dotted blue
histogram), and $e^+e^-\to\eta\gamma$ (dash-dotted green histogram)
at $\sqrt{s}=1282$ MeV. The arrow indicates the cut
$M_{\pi\gamma}<720$ MeV.
Right panel: The $Q_\gamma$ distribution for simulated events of the processes
$e^+e^-\to f_1 \to \eta\pi^0\pi^0$ (solid black histogram) and
$e^+e^-\to K_SK_L$ (dashed red histogram) at $\sqrt{s}=1282$ MeV. The
arrow indicates the cut $Q_\gamma<0$.
\label{fig3}}
\end{figure}

To reduce the background from the process $e^+e^-\to\pi^0\pi^0\gamma$,
a kinematic fit to the hypothesis $e^+e^-\to\pi^0\pi^0\gamma$ is
performed. During the fit, all possible five-photon subsets of the six 
reconstructed photons are tested. Events in which the $\chi^2$ value for one
of the combinations is less than 60 ($\chi^2_{\pi^0\pi^0\gamma}<60$) are
rejected. The simulated $\chi^2_{\pi^0\pi^0\gamma}$ distributions for
signal events and $e^+e^-\to\omega\pi^0\to\pi^0\pi^0\gamma$ background events
are shown in Fig.~\ref{fig2} (right). The condition
$\chi^2_{\pi^0\pi^0\gamma}>60$ reduces the
$e^+e^-\to\pi^0\pi^0\gamma$ background by a factor of more than 10, with a
signal loss of 20\%.

To calculate the next two parameters used for background suppression
in Ref.~\cite{SNDf1}, we use the photon energies and angles refined
after kinematic fitting to the $e^+e^-\to 6\gamma$ hypothesis. Most of
the events of the $e^+e^-\to\omega\pi^0(\gamma)$ process remaining
after applying the condition $\chi^2_{\pi^0\pi^0\gamma}>60$ have an
additional photon emitted from the initial state at a large angle. To
suppress this background, we require that the event not contain an
$\omega$-meson candidate, defined as a combination of three
photons: one is the most energetic photon in the event, and
the other two must have an invariant mass in the range
$|M_{2\gamma}-M_{\pi^0}|<35$ MeV. If there are several $\omega$-meson
candidates, the one with the smallest difference
$|M_{3\gamma}-M_{\omega}|$ is selected. The mass distributions of the
$\omega$-meson candidate $M_{\pi\gamma}$ for signal and
background events are shown in Fig.~\ref{fig3} (left).
Events with $M_{\pi\gamma}>720$ MeV are rejected. 

This condition
suppresses the background from the processes $e^+e^-\to\omega\pi^0$
and $e^+e^-\to\omega\pi^0\pi^0$ by approximately a factor of 15. For these
two processes, the $M_{\pi\gamma}$ distributions are similar. The
background from the process $e^+e^-\to \eta\gamma(\gamma)$ is
suppressed by more than a factor of 4. The efficiency for the process under
study is reduced by approximately a factor of 2. It is important to
note that for the $e^+e^-\to\pi^0\pi^0\gamma$ process with a uniform
phase-space distribution, the condition $M_{\pi\gamma}>720$ MeV
reduces the efficiency by a factor of only 2.5. This means that the calculated
background from the $e^+e^-\to\pi^0\pi^0\gamma$ process strongly
depends on the model used for its simulation. In the current analysis,
we use the simplest model, in which non-$\omega\pi^0$ events are simulated
assuming a uniform phase-space distribution, and the interference
between the $\omega\pi^0$ and non-$\omega\pi^0$ amplitudes is ignored.

The last condition applied in Ref.~\cite{SNDf1} is
$2E_{\gamma,\rm{max}}/\sqrt{s}<0.78$, where $E_{\gamma,\rm{max}}$ is the
energy of the most energetic photon in the event. After applying all
the previous conditions, it has practically no effect on the signal
detection efficiency and reduces the background from the processes
$e^+e^-\to \eta\gamma(\gamma)$ and $e^+e^-\to\omega\pi^0\pi^0$ by 30\%.

After applying the selection criteria from Ref.~\cite{SNDf1}, the
calculated background at 1.282 GeV contains 13\% of the events from the
$e^+e^-\to K_SK_L$ process. To suppress this background, the condition on 
the parameter $Q_\gamma=\max(L_{\gamma,5},L_{\gamma,6})$ is used, where
$L_{\gamma,i}$ is the negative logarithm of the probability that the
transverse energy profile in the calorimeter for the
$i$-th photon is obtained from a single photon. Photons 5 and 6 are
the photons that form the $\eta$-meson candidate. The $Q_\gamma$
distributions for signal and $e^+e^-\to K_SK_L$ background events
are shown in Fig.~\ref{fig3} (right). The
condition $Q_\gamma<0$ suppresses the background from the process
$e^+e^-\to K_SK_L$ by a factor of three.

We will refer to the set of selection conditions described above as the
tight selection. After applying the tight selection, 52
events are selected from the data. The energy distribution of these events
($N_1$) is presented in Table~\ref{tab2}.

\begin{table}
\caption{\label{tab2}
The c.m. energy, integrated luminosity ($L$), number of events that
have passed the tight ($N_1$) and loose ($N_2$) selection.}
\begin{ruledtabular}
\begin{tabular}{cccc}
$\sqrt{s}$ (GeV) &  $L$ (pb$^{-1}$) & $N_1$ & $N_2$ \\
\hline 
1.140&  4.8&   1&   4 \\
1.160&  9.9&   0&   6 \\
1.180&  4.9&   3&   4 \\
1.200&  6.5&   3&   8 \\
1.220&  5.8&   4&   5 \\
1.225&  2.6&   1&   3 \\
1.240&  5.9&   0&   5 \\
1.250&  1.0&   0&   2 \\
1.260& 10.4&   0&   7 \\
1.276&  9.5&   5&  21 \\
1.282& 62.7&  19&  90 \\
1.300&  7.7&   2&   8 \\
1.318& 10.9&   3&  12 \\
1.325&  1.1&   0&   2 \\
1.340&  5.3&   1&   7 \\
1.350&  2.6&   1&   3 \\
1.360& 11.0&   1&  14 \\
1.375&  1.0&   0&   1 \\
1.380&  4.8&   2&   8 \\
1.400&  7.2&   1&  18 \\
1.420&  5.3&   2&  12 \\
1.440&  5.8&   0&  18 \\
1.460& 12.0&   3&  30 \\
\end{tabular}
\end{ruledtabular}
\end{table}

\section{Fit of the energy distribution of the selected events}
\begin{figure}
\includegraphics[width=0.47\textwidth]{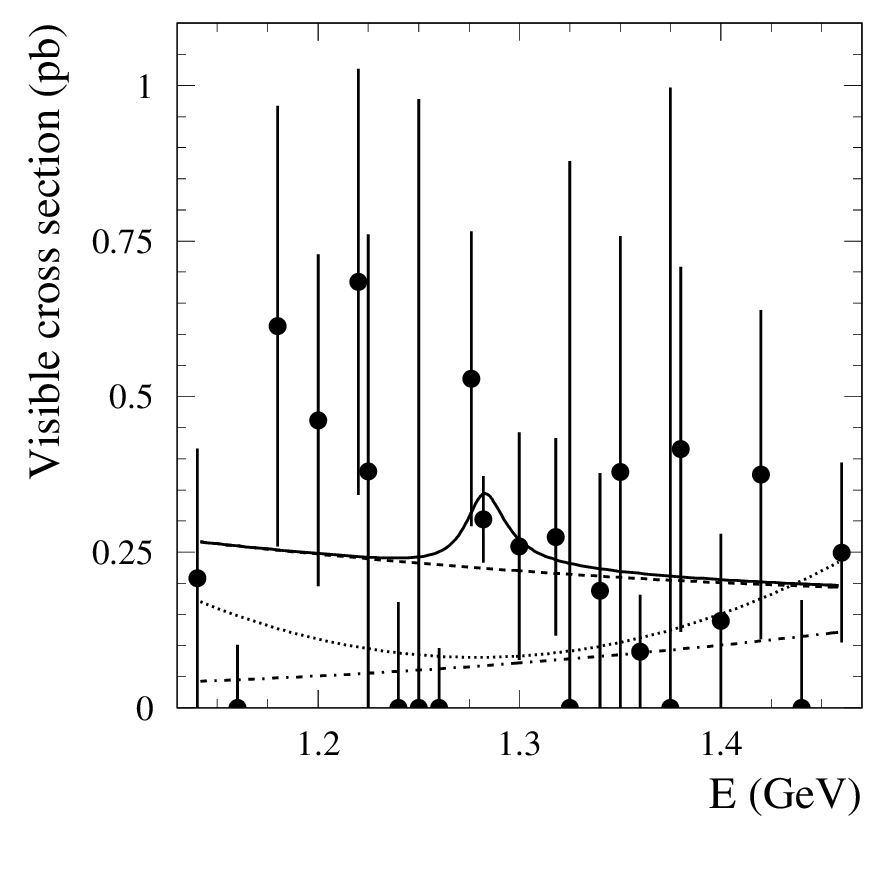}
\caption{The cross section for selected data events that pass the tight
selection (points with error bars). The dotted curve represents the 
total simulated background cross section, while the dash-dotted
curve represents the expected background from the processes
$e^+e^-\to\pi^0\pi^0\gamma$ and $e^+e^-\to\eta\pi^0\gamma$.
The solid curve is the result of the fit by the sum of the signal cross
section and the quadratic background. The dashed curve is the fitted
background cross section.
\label{fig4} }
\end{figure}

The observed cross section for the selected events is shown in
Fig.~\ref{fig4}. It is calculated as $N_1/L$, where $L$ is the integrated
luminosity. The cross section for background events obtained from
simulation is well described by a quadratic polynomial. This
polynomial is shown in Fig.~\ref{fig4} as a dotted curve. It is
evident that the simulation significantly underestimates the
background contribution. At $\sqrt{s}=1.282$ GeV, 
the simulation predicts the following proportions for the different
background processes:
$\pi^0\pi^0\gamma:\omega\pi^0\pi^0:\eta\gamma:
\eta\pi^0\gamma:K_SK_L=0.61:0.05:0.14:0.15:0.04$.
In the process $e^+e^-\to \pi^0\pi^0\gamma$, the fraction of events
from the non-$\omega\pi^0$ mechanism is 10\%. The selection criteria increase
this fraction by a factor of 10 compared to the simulation before
selection. We attribute the poor agreement in the background level between 
data and simulation to the inaccuracy of the model for
the process $e^+e^-\to \pi^0\pi^0\gamma$, as well as to a possible
difference between the data and the simulation in the
$\omega$-meson line shape in the processes $e^+e^-\to \omega\pi^0$
and $e^+e^-\to \omega\pi^0\pi^0$.

The distribution of the selected data events, given in Table~\ref{tab2}, is
fitted by the sum of the signal and background distributions:
\begin{equation}
N_i^{\rm th}=\varepsilon_i\sigma_{\rm vis}(\sqrt{s_i})L_i+
\sigma_{\rm bkg}(\sqrt{s_i})L_i,
\label{eq3}
\end{equation}
where $\varepsilon_i$ is the detection efficiency for the process
$e^+e^-\to f_1 \to \eta\pi^0\pi^0$, $\sigma_{\rm vis}(\sqrt{s_i})$ is the 
visible cross section for this process, and $\sigma_{\rm bkg}(\sqrt{s_i})$
is the background cross section at the energy point $\sqrt{s_i}$.

The visible cross section for the process $e^+e^-\to
f_1(1285)\to\eta\pi^0\pi^0$ is calculated as follows:
\begin{equation}
\sigma_{\rm vis}(\sqrt{s})=\int_{0}^{x_{\rm max}}
W(s,x)\sigma_{\rm B}(\sqrt{s(1-x)}) dx,
\label{eq1}
\end{equation}
where $W(s,x)$ is a radiator function that describes the probability
of emitting photons with total energy $x\sqrt{s}/2$ from the initial
state~\cite{rad1}. The Born cross section is parametrized as follows:
\begin{equation}
\sigma_{\rm B}(\sqrt{s})=\sigma(e^+e^-\to f_1)B(f_1(1285)\to
\eta\pi^0\pi^0)
\frac{m^2_{f_1}\Gamma^2}{(s-m^2_{f_1})^2+m^2_{f_1}\Gamma^2}
\frac{m^3_{f_1}P(s)^3}{s^{3/2}P(m^2_{f_1})^3},
\label{eq2}
\end{equation}
where $\sigma(e^+e^-\to f_1)=(12\pi/m^2_{f_1})B(f_1\to e^+e^-)$ is the cross
section at the resonance maximum.
In Eq.~(\ref{eq2}), we assume that the $f_1\to \eta\pi^0\pi^0$ decay
proceeds via the $a_0(980)\pi^0$ intermediate state. Therefore, $P(s)$
is the momentum of the $a_0(980)$ meson. Radiative corrections reduce
the visible cross section at the resonance maximum by 20\% compared to
the Born cross section.

The detection efficiency of $e^+e^-\to \eta\pi^0\pi^0$ events
is calculated using simulation under the assumption that the
$f_1(1285)$ decay to this final state proceeds via the intermediate mechanisms
$a_0^0(980)\pi^0$ and $f_0(500)\eta$ in the proportion
0.72:0.28~\cite{pdg}. The efficiency varies from $(6.1\pm 0.1)\%$ at
$\sqrt{s}=1.2$ GeV to $(5.0\pm 0.4)\%$ at $\sqrt{s}=1.282$ GeV and to
$(1.8\pm 0.2)\%$ at $\sqrt{s}=1.4$ GeV, where the quoted errors represent the
model uncertainty, estimated as the difference in efficiencies in the
$a_0^0(980)\pi^0$ and $a_0^0(980)\pi^0 + f_0(500)\eta$ models. A
detailed study of the systematic uncertainty associated with the
selection of multiphoton events based on the kinematic fit was
performed in Refs.~\cite{SNDomegapi} for events of the
$e^+e^-\to \pi^0\pi^0\gamma$ process. We estimate that the systematic
uncertainty in the detection efficiency due to inaccuracy in the
detector simulation does not exceed 5\%.

The background cross is described by a quadratic polynomial.
The free fit parameters are $\sigma(e^+e^-\to f_1)$ and the three
coefficients of the quadratic polynomial.

The fitting curve is shown in Fig.~\ref{fig4}. From the fit,
the following value of the cross section at the resonance maximum is
obtained:
\begin{equation}
\sigma(e^+e^-\to f_1)=14\pm16\mbox{ pb}.
\label{result}
\end{equation} 
A goodness-of-fit test is performed using an ensemble of toy
experiments, in which the Poisson-distributed data are generated
according to the fitted function.
These simulated data are fitted, and the distribution of the function
$\chi^2_{\lambda,p}$~\cite{baker} is calculated.
The experimental value of this function corresponds to a $p$-value
of 0.1.

A shortcoming of the fit described above is that the fitted background
energy dependence is nearly linear with a negative slope, contrary to
expectations. The energy
dependence of the expected background is shown in Fig.~\ref{fig4}. 
The rise of the background near the edges of the energy range under
study is due to radiative return to the $\phi(1020)$ resonance (via $e^+e^-\to
K_SK_L(\gamma)$ and $\eta\gamma(\gamma)$) at the left edge, and the
processes $e^+e^-\to\omega\pi^0\pi^0$ and $e^+e^-\to \omega\eta$
at the right edge. The dash-dotted curve in Fig.~\ref{fig4}
represents the expected background from the processes 
$e^+e^-\to\pi^0\pi^0\gamma$ and $e^+e^-\to\eta\pi^0\gamma$. 
The cross section for both processes increases with energy. Therefore,
we expect a positive slope for the background cross section near the
center of the energy range, with a steep rise near its
edges. The incorrect background description can lead to a biased
determination of the signal magnitude in the fit.

Loosening the selection conditions may enable a more accurate
determination of the background shape. Specifically, if 
we remove the conditions $M_{\pi\gamma}>720$ MeV and
$2E_{\gamma,\rm{max}}/\sqrt{s}<0.78$, the number of selected events
increases by a factor of 5. The signal detection efficiency increases 
approximately by a factor of 2 [up to $(9.1\pm 0.1)\%$
at $\sqrt{s}=1.2$ GeV, $(9.9\pm0.4)\%$ at $\sqrt{s}=1.282$ GeV, and
$(6.9\pm 0.2)\%$ at $\sqrt{s}=1.4$ GeV]. 
Therefore, the signal sensitivity, proportional to 
$S/\sqrt{S+B}$, where
$S$ and $B$ are the number of signal and background events
passing the selection criteria, does not deteriorate significantly.
We also expect a reduction in
the model dependence of the calculated background from
the $e^+e^-\to \pi^0\pi^0\gamma$ process, as well as a smaller data--simulation
difference related to the $\omega$-meson line shape.

\begin{figure}
\includegraphics[width=0.47\textwidth]{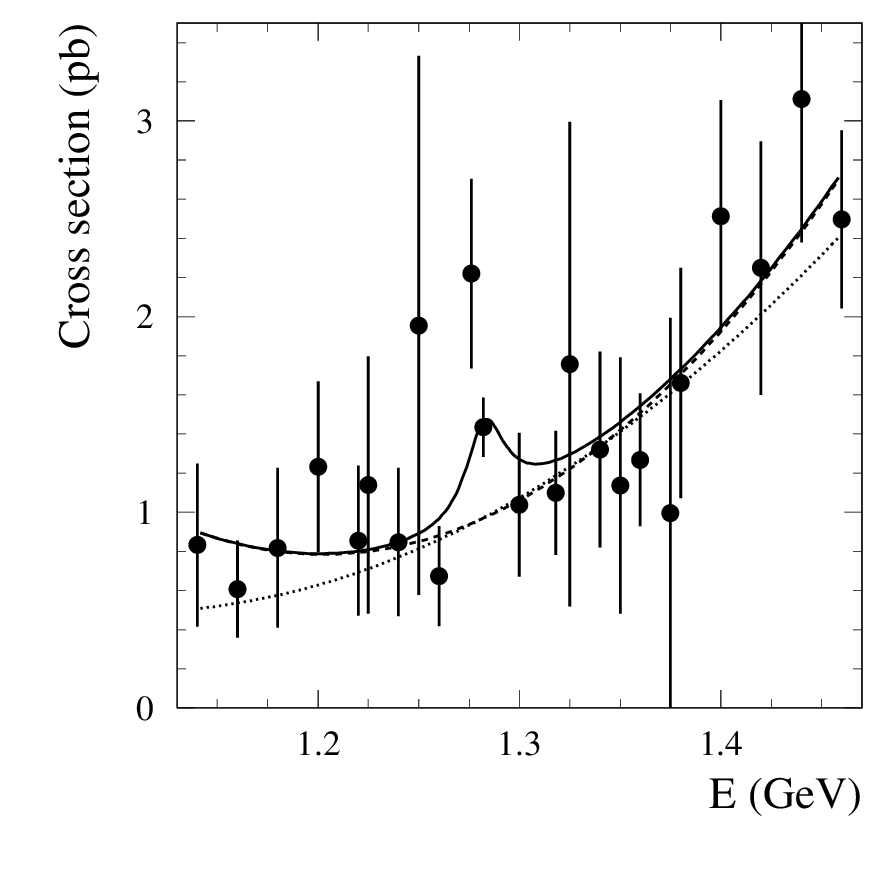}
\includegraphics[width=0.47\textwidth]{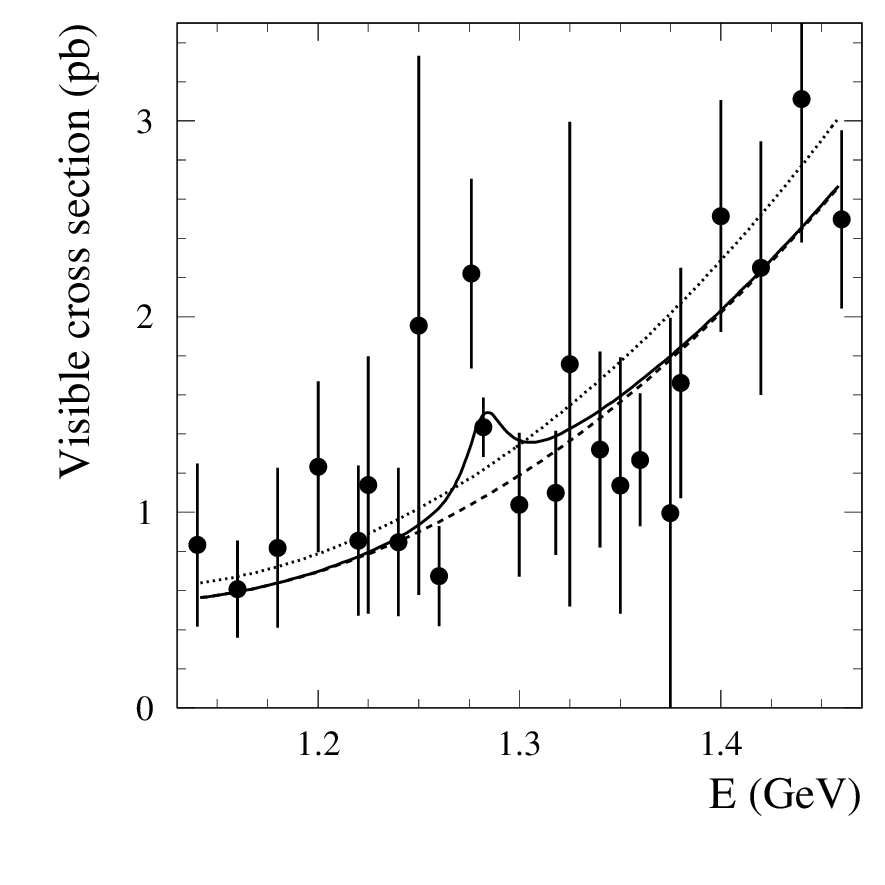}
\caption{The cross section for data events selected with loosened
conditions (points with error bars). In the left panel, the dotted curve
represents the background cross section calculated using the simulation, the
solid curve is the result of the fit to the data with a quadratic background,
and the dashed curve is the fitted background cross section. 
In the right
panel, the solid curve is the result of the fit using the scaled
simulated background, the dashed curve is the fitted background cross
section, and the dotted curve is the result of the fit without
the resonance.
\label{fig5}}
\end{figure}

The cross section for events selected with loosened conditions is
shown in Fig.~\ref{fig5}. The left plot shows the result of the fit to the
cross section using Eq.~(\ref{eq3}) with a quadratic background. The
$p$-value for this fit is 0.85. The fitted cross section at the
resonance maximum is
\begin{equation}
\sigma(e^+e^-\to f_1)=40\pm18\mbox{ pb}.
\label{result1}
\end{equation} 
Compared to the result~(\ref{result}) obtained for events selected
with the tight conditions, the statistical error increased by 12\%.

The dotted curve in Fig.~\ref{fig5} (left) shows the calculated
background cross section. It is evident that with the loosened
selection the simulation reproduces both the shape and the level of
the background reasonably well. Simulation predicts the following proportions
for different background processes at $\sqrt{s}=1.282$:
$\pi^0\pi^0\gamma:\omega\pi^0\pi^0:\eta\gamma:\eta\pi^0\gamma:K_SK_L=
0.77:0.11:0.08:0.03:0.02$.
Figure~\ref{fig5} (right) shows the result of the fit with
Eq.~(\ref{eq3}), where the simulated background cross section
is multiplied by a factor $\alpha_{\rm bkg}$. The free
parameters of the fit are $\alpha_{\rm bkg}$ and $\sigma(e^+e^-\to
f_1)$. The $p$-value for this fit is 0.83. The fitted cross section
at the resonance maximum is
\begin{equation}
\sigma(e^+e^-\to f_1)=31\pm13\pm4\mbox{ pb}.
\label{result2-1}
\end{equation}
The first error of the cross section is statistical, and the second is
systematic. The systematic error includes the model (4\%) and systematic (5\%)
uncertainties in the detection efficiency, the uncertainty in the 
$f_1\to\eta\pi^0\pi^0$ branching fraction (4\%), the uncertainty in the 
luminosity measurement (1\%), and the model error associated with the 
background shape (11\%). 

The fitted scale factor is $\alpha_{\rm bkg}=1.11\pm0.09\pm0.10$, where the
first error is statistical, and the second is the model uncertainty associated
with the background shape.

The model uncertainty associated with the background shape is estimated
by varying the scale factors for background components within their
uncertainties. For the processes $e^+e^-\to \eta\gamma(\gamma)$ and $e^+e^-\to
\pi^0\pi^0\gamma$, the cross section is known to a few percent accuracy. 
To estimate the uncertainty in the scale factor for the main background 
process $e^+e^-\to \pi^0\pi^0\gamma$, we analyse the control region 
$25<\chi^2_{\pi\pi\gamma}<60$ enriched in $e^+e^-\to\pi^0\pi^0\gamma$ events.
A fit with the scale factor for background and null signal describes the data
well ($p$-value=0.87) and yields 
$\alpha_{\rm bkg}^{\pi\pi\gamma}=0.96\pm0.06$.
We conservatively estimate the systematic uncertainty in the
scale factor due to imperfect simulation of the detection efficiency
for the $e^+e^-\to \pi^0\pi^0\gamma$ process to be 10\%.
The same value for the uncertainty is used for the $e^+e^-\to
\eta\gamma(\gamma)$ process, for which the detection efficiency is
expected to be better simulated than for $e^+e^-\to \pi^0\pi^0\gamma$.
For the processes $e^+e^-\to \eta\pi^0\gamma$ and
$e^+e^-\to \omega\pi^0\pi^0$, the cross sections in the energy region
under study are known with an accuracy of about 30\%. For the process 
$e^+e^-\to K_SK_L(\gamma)$, the uncertainty comes from the
data/simulation difference in the detection efficiency, which is
estimated as 75\% from the study of the control region 
$35<\chi^2_{\eta\pi\pi}<70$. The shifts in $\sigma(e^+e^-\to f_1)$ and
$\alpha_{\rm bkg}$ obtained for variations of different background 
components are combined in quadrature.

The accuracy of the background calculation and model error estimation 
is tested in the control region $35<\chi^2_{\eta\pi\pi}<70$. In this region 
there is a 15\% contribution from the process $e^+e^-\to K_SK_L(\gamma)$,
which is poorly simulated and may worsen the agreement between data and
simulation. The scale factor obtained with the background-only
fit is $\alpha_{\rm bkg}^{\rm CR}=1.17\pm0.08\pm0.15$. The $p$-value for the
fit is 0.68. The model uncertainty in
$\alpha_{\rm bkg}^{\rm CR}$ is calculated as described above, and
is dominated by the uncertainty of the scale factor for the 
$e^+e^-\to K_SK_L(\gamma)$ process. 

Since the fitted value of the scale factor for the simulated
background is consistent with expectation, and the quality of the fit
with the scale factor is no worse than that with a quadratic
background, we adopt the result (\ref{result2-1}) as the main result.
Using information about the background shape resulted in a reduction
in the statistical uncertainty compared to (\ref{result1}). 

The statistical significance of the $e^+e^-\to f_1(1285)$
signal is estimated from the difference in the likelihood functions for 
the hypotheses with and without the resonance, and is $2.5\sigma$. It should
be noted that the $p$-value for the fit without resonance (0.55) is worse than 
that for the fit with resonance, though still reasonably good. However, the 
scale factor $\alpha_{\rm bkg}^{\rm nosig}=1.26\pm0.07\pm 0.10$
differs from unity by $2.1\sigma$. The result of the fit without resonance 
is shown in Fig.~\ref{fig5} (right). 
To estimate how the uncertainty in the background shape decreases the
signal significance, we include the scale factors for the background
components, together with the uncertainties defined above, in the fit
as nuisance parameters. A normal distribution is assumed for the
nuisance parameters. The significance of the
$e^+e^-\to f_1(1285)$ signal decreases to $2.4\sigma$.

The obtained value of the cross section at the resonance maximum
(\ref{result2-1}) corresponds to the branching fraction
\begin{equation}
B(f_1(1285)\to e^+e^-)=(3.5\pm 1.4\pm0.5)\times 10^{-9}. 
\label{result2-2}
\end{equation}
This value agrees with the theoretical predictions~\cite{Rudenko,HKZ}.
Our result is more than a factor of 2 lower than the CMD-3 value 
$(8.3\pm 2.3\pm1.1)\times10^{-9}$~\cite{cmdf1}. However, due to large 
uncertainties, the significance of this difference is only $1.6\sigma$. 
It should be noted that the current version (v2) of the CMD-3 
preprint~\cite{cmdf1} did not account for the nonresonant background,
in particular that from the $e^+e^-\to \eta\pi^0\gamma$ process.

Since the significance of the $f_1(1285)$ signal is low, we
also provide upper limits at the 90\% confidence level:
\begin{equation}
\sigma(e^+e^-\to f_1)<48\mbox{ pb, }
B(f_1(1285)\to e^+e^-)<5.4\times 10^{-9}.
\label{result2-3}
\end{equation}
The limit is calculated assuming a Gaussian distribution, as the central 
value plus 1.28 times the combined statistical and systematic
uncertainties.

\section{Summary}
In this paper, we present the results of a search for the production
of a single $f_1(1285)$ meson in $e^+e^-$ collisions. The analysis has
been performed using data with an integrated luminosity of 200
pb$^{-1}$, accumulated in the SND experiment at the VEPP-2000 $e^+e^-$
collider in the center-of-mass energy range of 1.14--1.46 GeV. About
72 pb$^{-1}$ were recorded at two points (1.276 and 1.282 GeV) near the
maximum of the $f_1(1285)$ resonance. To search for the $e^+e^-\to
f_1(1285)$ process, the $f_1(1285)$-meson decay channel to
$\eta\pi^0\pi^0$ with subsequent decay of $\eta$ and $\pi^0$ mesons to
$\gamma\gamma$ has been used. From the fit to the energy dependence of
the cross section for selected $\eta\pi^0\pi^0$ candidates by the sum
of the signal and background distributions, the cross section at the
resonance maximum $\sigma(e^+e^-\to f_1)=(31\pm 13\pm 4)$ pb and the
branching fraction 
$B(f_1(1285)\to e^+e^-)=(3.5\pm 1.4\pm 0.5)\times 10^{-9}$ have been
obtained. 
The significance of the $e^+e^-\to f_1(1285)$
signal is $2.4\sigma$. The measured  $B(f_1(1285)\to e^+e^-)$ is
consistent with the theoretical predictions~\cite{Rudenko,HKZ}. Our result
is $1.6\sigma$ lower than the CMD-3 measurement
$B(f_1(1285)\to e^+e^-)=(8.3\pm 2.3\pm1.1)\times 10^{-9}$~\cite{cmdf1}.

Since the significance of the $f_1(1285)$ signal is low, we also
provide upper limits at the 90\% confidence level: $\sigma(e^+e^-\to
f_1)<48\mbox{ pb}$ and
$B(f_1(1285)\to e^+e^-)<5.4\times 10^{-9}$. The results obtained in
this work supersede those of Ref.~\cite{SNDf1}.

\end{document}